\documentclass[lettersize,journal]{IEEEtran}
\usepackage{amsmath,amsfonts}
\usepackage{algorithm}
\usepackage{array}
\usepackage[caption=false,font=normalsize,labelfont=sf,textfont=sf]{subfig}
\usepackage{textcomp}
\usepackage{stfloats}
\usepackage{url}
\usepackage{verbatim}
\usepackage{graphicx}
\usepackage{cite}
\usepackage{enumitem}
\usepackage{multirow}%
\usepackage{colortbl}  

\usepackage{booktabs}
\usepackage{xcolor}
\newcommand{\Xhline}[1]{%
  \noalign{\ifnum0=`}\fi\hrule height #1%
  \futurelet\reserved@a\@xhline}
\usepackage{algpseudocode}
\usepackage[normalem]{ulem}  
\usepackage{amsthm}
\newtheoremstyle{boldlemma}
  {\topsep}   
  {\topsep}   
  {\itshape}  
  {}          
  {\bfseries} 
  {.}         
  {.5em}      
  {}          
\theoremstyle{boldlemma}
\newtheorem{lemma}{Lemma}[section]
\usepackage{hyperref}
    \hypersetup{
        bookmarksnumbered=true,
        bookmarksopen=true,
        bookmarksopenlevel=5,
        colorlinks=true,
        citecolor=blue,
        filecolor=black,
        linkcolor=black,
        urlcolor=black,
        pdfstartview=Fit,
        pdfpagemode=UseOutlines
    }
\begin{document}


\title{Automatic Phase Calibration for High-resolution mmWave Sensing via Ambient Radio Anchors}


\author{Ruixu~Geng,
        Yadong~Li,
        Dongheng~Zhang,~\IEEEmembership{Member,~IEEE},
        Pengcheng~Huang,
        Binquan~Wang,
        Binbin~Zhang,
        Zhi~Lu,~\IEEEmembership{Member,~IEEE},
        Yang~Hu,~\IEEEmembership{Member,~IEEE},
        Yan~Chen,~\IEEEmembership{Senior~Member,~IEEE}
        
        \thanks{This work has been submitted to the IEEE for possible publication. Copyright may be transferred without notice, after which this version may no longer be accessible.
        }
        \thanks{Ruixu Geng, Dongheng Zhang, Pengcheng Huang, Binquan Wang, Binbin Zhang, Zhi Lu, and Yan Chen are with the School of Cyber Science and Technology, University of Science and Technology of China, Hefei 230026, China (e-mail: \{gengruixu, wzwyyx, lijiamu, chenqi777\}@mail.ustc.edu.cn, \{dongheng, eecyan\}@ustc.edu.cn)}
        \thanks{Yang Hu is with the School of Information Science and Technology, University of Science and Technology of China, Hefei 230026, China (e-mail: eeyhu@ustc.edu.cn)}
        \thanks{Yadong Li is with the Department of Electrical \& Computer Engineering, University of Washington, Seattle, WA 98195, USA (e-mail: yadongli@uw.edu)}
        \thanks{This paper has supplementary downloadable material available at https://ieeexplore.ieee.org, provided by the authors. The material includes a pdf document supplementing some details of the paper. Contact eecyan@ustc.edu.cn for further questions about this work.}
}

\markboth{IEEE Transactions on Mobile Computing,~Vol.~XX, No.~X, July~2025}%
{Geng \MakeLowercase{\textit{et al.}}: Automatic Phase Calibration for High-resolution mmWave Sensing via Ambient Radio Anchors}


\maketitle

\begin{abstract}
Millimeter-wave (mmWave) radar systems with large array have pushed radar sensing into a new era, thanks to their high angular resolution. However, our long-term experiments indicate that array elements exhibit phase drift over time and require periodic phase calibration to maintain high-resolution, creating an obstacle for practical high-resolution mmWave sensing. Unfortunately, existing calibration methods are inadequate for periodic recalibration, either because they rely on artificial references or fail to provide sufficient precision.
To address this challenge, we introduce AutoCalib, the first framework designed to automatically and accurately calibrate high-resolution mmWave radars by identifying Ambient Radio Anchors (ARAs)—naturally existing objects in ambient environments that offer stable phase references. AutoCalib achieves calibration by first generating spatial spectrum templates based on theoretical electromagnetic characteristics. It then employs a pattern-matching and scoring mechanism to accurately detect these anchors and select the optimal one for calibration.
Extensive experiments across 11 environments demonstrate that AutoCalib capable of identifying ARAs that existing methods miss due to their focus on strong reflectors. AutoCalib's calibration performance approaches corner reflectors (74\% phase error reduction) while outperforming existing methods by 83\%. Beyond radar calibration, AutoCalib effectively supports other phase-dependent applications like handheld imaging, delivering 96\% of corner reflector calibration performance without artificial references.
\end{abstract}


\section{Introduction} \label{sec:intro}

Millimeter-wave (mmWave) radar with large antenna arrays opens exciting possibilities for precise sensing and imaging, but fully harnessing its power requires accurately capturing subtle phase shifts between antennas.
There is a widely accepted experience in mmWave MIMO radar sensing: the radar antenna arrays typically require only one-shot calibration—once calibrated using corner reflectors in anechoic chambers, the antenna arrays can maintain their phase over long time periods~\cite{delafrooznorooziZeroShotAccurateMmWave2024}. This experience has led most deployments, from autonomous vehicles~\cite{tagliaferriCooperativeCoherentMultistatic2024,yaoRadarPerceptionAutonomous2023,huangOverviewSignalProcessing2023,zhuMaliciousAttacksMultiSensor2024} to human sensing applications~\cite{zhangSurveyMmWaveBasedHuman2023,zhangMonitoringLongtermCardiac2024,shastriReviewMillimeterWave2022,huContactlessArterialBlood2024}, to operate without periodic (e.g., monthly) phase recalibration.

\begin{figure}[!tbp]
    \centering
    \includegraphics[width=0.94\linewidth]{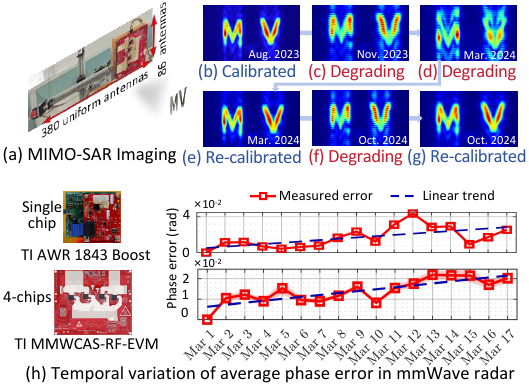}
    \caption{Phase error between mmWave radar antennas changes over time. Long-term experiments show that this change is fatal for MIMO-SAR imaging applications using TI cascade radar (N=86).}
    \label{fig1:phase_error}
\end{figure}

However, our long-term experiments challenge this assumption. In August 2023, we calibrated a TI 4-chips cascaded radar for a MIMO-SAR imaging experiment~\cite{wangPrecise2D3D2023,albabaSidelobesGhostTargets2024,liHighresolutionHandheldMillimeterwave2024} in which the radar was positioned vertically and moved horizontally using a motion stage (Fig.~\ref{fig1:phase_error}(a)). The system, comprising 86 MIMO antennas arranged as a vertical array, yielded exceptionally high imaging quality (Fig.~\ref{fig1:phase_error}(b)). Three months later, we observed that the image quality in the vertical direction blurred (Fig.~\ref{fig1:phase_error}(c)). Another four months later, the vertical blurring had become severe and greatly affected the overall imaging quality (Fig.~\ref{fig1:phase_error}(d)). A quick recalibration restored high quality (Fig.~\ref{fig1:phase_error}(e)), but the imaging results had again degraded noticeably seven months later (Fig.~\ref{fig1:phase_error}(f)), and improved after another recalibration (Fig.~\ref{fig1:phase_error}(g)). \textit{These observations point to continuous phase drift that contradicts the one-shot calibration assumption.}
Complementary phase error measurements on both a TI 4-chips radar (86 antennas) and a TI single chip radar~\cite{instrumentsAWR1843AutomotiveRadar2018} (8 antennas) over 14 days with over 50 measurements per day confirmed clear and consistent variations in the inter-antenna phase responses, as shown in Fig.~\ref{fig1:phase_error}(h). On average, each antenna exhibited a phase drift of 0.0005 rad per day.\footnote{See Supplementary Note 1 and https://tmc2025.vercel.app/ for details}

\begin{figure}[!tbp]
    \centering
    \includegraphics[width=0.94\linewidth]{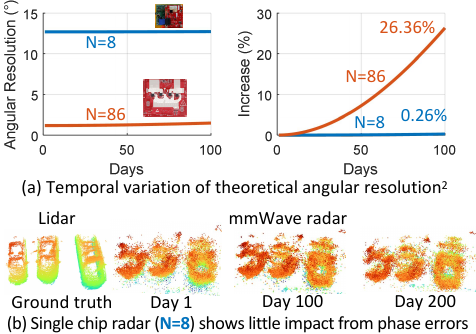}
    \caption{(a) Impact of phase drift on different radars. (b) TI 1843 Boost radar (N=8) shows little impact.}
    \label{fig2:phase_impact}
\end{figure}

At first sight, such modest phase variations may seem inconsequential. As shown in Fig.~\ref{fig2:phase_impact}(a), commonly used single-chip radar (8 antenna arrays) experience only little degradation in angular resolution: merely 0.26\% after 100 days without calibration.\footnote{See Supplementary Note II for details} This has little impact on tasks such as respiratory monitoring~\cite{wangRFGymCareIntroducingRespiratory2024,bauderMMMUREMmWaveBasedMultiSubject2024}, pose estimation~\cite{caoMmCLIPBoostingMmWavebased2024,wengLargeModelSmall2024}, or point cloud reconstruction~\cite{chengNovelRadarPoint2022,gengDREAMPCDDeepReconstruction2024,caiMilliPCDTraditionalVision2023}. Fig.~\ref{fig2:phase_impact}(b) confirms this tolerance, showing negligible point cloud degradation in a single-chip radar even after 200 days. This resilience has caused the mmWave radar sensing community to overlook the calibration challenge.
However, as manufacturing processes improve and array sizes increase~\cite{karimian-sichaniAntennaArrayWaveform2024,turkmen223276GHzCascadable2023}, periodic calibration becomes critical. Fig.~\ref{fig2:phase_impact}(a) clearly illustrates that the 4-chip mmWave radar (N=86) could experience up to a 26.36\% loss in angular resolution over just 100 days, which explains the image degradation observed in Fig.~\ref{fig1:phase_error}(a). As future mmWave radar systems incorporate even more antennas for higher angular resolution~\cite{zhengEnhancingMmWaveRadar2024}, this problem will become increasingly severe. Tab.~S1 in Supplementary Note II demonstrates that after 100 days, arrays with 200 and 400 antennas would suffer 111.9\% and 286.8\% degradation in angular resolution, respectively. Thus, high-resolution mmWave radar systems require periodic calibration to maintain angular resolution, creating an urgent need for reliable and automatic calibration methods.

Existing calibration techniques broadly fall into three categories~\cite{tsaregorodtsevAutomatedAutomotiveRadar2023,heiningOvertheAirCalibrationMmW2023}. Mainstream methods use metal reflectors with known shapes, such as corner reflectors~\cite{xu3DHighResolutionImaging2024,raphaeliChallengesAutomotiveMIMO2023}, metal plates~\cite{sprengUWBNearfieldMIMO2013}, and metal spheres~\cite{liuNearFieldCalibrationMillimeterWave2024,guoMillimeterWave3DImaging2020}. Although these methods offer high precision, they require the manual placement of specialized reflectors and professional calibration procedures. For mmWave radars in consumer applications (such as smart vehicles/smart homes), it is difficult for ordinary users to perform regular calibration independently.
Other solutions utilize strong scatterers to exploit naturally occurring reference~\cite{zhangMobi2Sense2022,zhangRFSearchSearchingUnconscious2023,laiEnablingVisualRecognition2024, tagliaferriCooperativeCoherentMultistatic2024}. However, their precision is limited: Unlike corner reflectors that provide predictable triple bounces, most natural scatterers typically produce complex reflections that vary with slight antenna changes (i.e., lack of stable phase center), making them 
unreliable for mmWave precise phase calibration~\cite{xu3DHighResolutionImaging2024}.
Additionally, researchers have attempted self-calibration using antenna coupling~\cite{tianPragmaticApproachPhase2021,yuAdaptivePhasearrayCalibration2013,aumannPhasedArrayAntenna1989} to eliminate the need for reflectors, but this approach is impractical for commercial mmWave radar systems due to their intentionally minimized inter-antenna coupling~\cite{texasinstrumentsTIMmWaveRadar2018} and restricted Field-of-View (FOV)~\cite{texasinstrumentsUsersGuideAWR1843BOOST2020}. In summary, there is currently no fully automatic and precise calibration solution for commercial mmWave radars. This gap motivates us to question a fundamental assumption: \textit{why do corner reflectors provide precise calibration while most naturally occurring strong reflectors fall short?}

\noindent \textbf{Insight.}
Corner reflectors succeed primarily because they satisfy a crucial condition: \uline{their radar cross-section (RCS) remains consistent across all receiving antennas in the array.} This stable phase center property, rather than merely strong reflectivity, makes them ideal for calibration. Our insight is that certain objects in the ambient environment naturally possess such similar characteristics and can be used as anchors for calibration. We call these objects \uline{Ambient Radio Anchors (ARAs)}.

While corner reflectors achieve the stable phase property through specialized geometric design, classic electromagnetic theory~\cite{hansenLightScatteringPlanetary1974,balanisAdvancedEngineeringElectromagnetics2012} suggests small objects can also function as ARAs when their dimensions approach the wavelength ($\approx$4mm), as their scattering becomes dominated by resonance effects rather than geometric optics. This means small metallic objects like screws, rivets, and electronic components can naturally function as ARAs, creating stable phase centers despite having much weaker reflections than deliberately engineered reflectors

Our preliminary experiments in Fig.~\ref{fig3:calibration_comparison} (details in Sec.~\ref{sec:ExistingCalibrationApproaches}) validated that an isolated ARA (a screw) can achieve calibration precision comparable to corner reflector and superior to ordinary strong reflectors. The critical problem lies in accurately identifying these ARAs within complex environments.

\noindent \textbf{Challenges and opportunities.}
However, there exist two challenges: (1) Different objects exhibit varied reflection characteristics, making it difficult to identify those with properties closest to ideal ARAs among diverse environmental reflectors. (2) Environmental characteristics vary, and a robust detection method need work across diverse settings. Fortunately, by leveraging the mathematical definition of ARAs, we have derived two key properties: (1) Virtual Isolated Point (VIP), which means the reflection characteristics of an ARA are highly consistent with those of an ideal isolated point target. If a measured reflection pattern closely matches the VIP profile, it can reliably be identified as an ARA. (2) Pattern Invariance (PI), which indicates that the reflection signature of an ARA remains stable across different environments, showing only global phase shifts without changing its structural characteristics. These properties provide a practical solution for ARA detection: as long as the measured pattern matches the expected template of an isolated point, the target can be consistently detected regardless of environmental changes.

\noindent \textbf{Our approach.} 
To transform these theoretical insights into a practical solution, we present AutoCalib, a three-step system that automatically discovers and validates ARAs for mmWave calibration. First, we build a template library by selecting spatial spectrum templates that capture the ideal ARA characteristics derived from VIP and PI properties. Second, we perform pattern matching by transforming radar measurements into spatial spectrum representations and matching them against the theoretical templates to identify candidate ARAs. Third, we rank the detected candidates using a comprehensive scoring model that favors targets with stable phase centers. This systematic design enables AutoCalib to support a wide range of applications including radar array calibration~\cite{raphaeliChallengesAutomotiveMIMO2023}, handheld imaging~\cite{liHighresolutionHandheldMillimeterwave2024,liIFNetDeepImaging2024}, and human sensing~\cite{zhangMobi2Sense2022,maMobi2SenseEnablingWireless2022}.

Extensive experiments across 11 diverse environments demonstrate that our system accurately identifies pre-placed ARAs and discovers numerous hidden ARAs that were previously undetectable. When used for mmWave radar calibration, these ARAs achieve performance comparable to professional corner reflectors while outperforming existing methods by 83\%. Micro-benchmark experiments further validate our system's robust performance against variations in distance, direction, angle, and phase errors. In practical applications requiring phase error estimation, such as handheld imaging, our approach achieves 96\% of the performance of corner reflector-based methods without requiring any artificial reference objects, enabling truly automated calibration in everyday environments.

AutoCalib makes the following contributions:
\begin{enumerate}[leftmargin=*,topsep=0.6em]  
    \item For the first time, we quantify the periodically calibration requirement for mmWave radar systems, demonstrating through rigorous experiments that phase stability degrades more than previously recognized, particularly for large antenna arrays.

    \item We introduce the concept of Ambient Radio Anchors (ARAs), and propose AutoCalib, a novel system that enables automatic and precise calibration through the discovery and validation of ARAs in everyday environments, eliminating the need for artificial reflectors or expert intervention.

    \item We comprehensively evaluate AutoCalib across diverse sensing applications and environments, demonstrating calibration quality comparable to professional methods while maintaining robustness to environmental variations and viewing angles.
\end{enumerate}

\section{Preliminary}
This section first reviews existing mmWave radar calibration methods, analyzes their limitations, and then reveals the opportunity of using Ambient Radio Anchors (ARAs) for automatic and accurate mmWave radar calibration.

\subsection{Phase Errors in mmWave Radar}

In practical mmWave radar systems, an N-element antenna array operates in an environment containing a distribution of spatial scatterers. Let $\Omega$ represent this scattering environment. Under monostatic antenna assumption, the ideal received signal at the $n$-th antenna position $\mathbf{p}_n$ can be modeled using an integral over all scattering points $\mathbf{p}_\omega$ in $\Omega$:

\begin{equation}
y_n^{\text{ideal}} = \int\limits_{\omega \in \Omega} \gamma(|\mathbf{p}_n-\mathbf{p}_\omega|) \sigma(\mathbf{p}_\omega; \mathbf{p}_n) e^{j\Phi(|\mathbf{p}_n-\mathbf{p}_\omega|)} \, d\mathbf{p}_\omega,
\label{eq:ideal_signal}
\end{equation}
where $\gamma(|\mathbf{p}_n-\mathbf{p}_\omega|)$ represents propagation attenuation over distance $|\mathbf{p}_n-\mathbf{p}_\omega|$, $\sigma(\mathbf{p}_\omega; \mathbf{p}_n)$ is the RCS at point $\mathbf{p}_\omega$ observed from position $\mathbf{p}_n$, $\Phi(|\mathbf{p}_n-\mathbf{p}_\omega|) = 2\pi f_c |\mathbf{p}_n-\mathbf{p}_\omega|/c$ is the phase delay determined by round-trip propagation time, and $f_c$ is the carrier frequency.

In practice, each antenna introduces phase errors, yielding:

\begin{equation}
y_n^{\text{actual}} = \int\limits_{\omega \in \Omega} \gamma(|\mathbf{p}_n-\mathbf{p}_\omega|) \sigma(\mathbf{p}_\omega; \mathbf{p}_n) e^{j(\Phi(|\mathbf{p}_n-\mathbf{p}_\omega|) + \Delta\phi_n)} \, d\mathbf{p}_\omega,
\label{eq:actual_signal}
\end{equation}
where $\Delta\phi_n$ represents the phase error from manufacturing tolerances, temperature drift, oxidation, and synchronization issues~\cite{texasinstrumentsApplicationNoteCascade2022}. Despite commercial radars' embedded calibration routines~\cite{texasinstrumentsUsersGuideAWRx2020}, residual errors accumulate over time.

\subsection{Existing Calibration Approaches} \label{sec:ExistingCalibrationApproaches}

The goal of mmWave radar phase calibration is to accurately estimate the phase error $\Delta \phi_n$ for each antenna to restore the original phase coherence relationships. 

\subsubsection{Calibration Using Artificial Reflectors} \label{sec:calibration_reflectors}

In this method, reflectors with precisely known dimensions and radio behavior are deliberately placed in the radar's field of view.

Corner reflectors are particularly useful because they can be modeled as point targets with stable phase centers~\cite{sarabandiOptimumCornerReflectors1996,xu3DHighResolutionImaging2024,raphaeliChallengesAutomotiveMIMO2023}. Specifically, their scattering behavior satisfies the condition:
\begin{equation}
\frac{\partial \sigma(\mathbf{p}_\omega; \mathbf{p}_n)}{\partial \mathbf{p}_n} = 0, \forall \omega \in \Omega,
\label{eq:corner_reflector_condition}
\end{equation}
meaning their radar response is consistent regardless of viewing angle. If a corner reflector is positioned at location $\mathbf{p}_0$, its contribution to the ideal received signal can be simplified to:
\begin{equation}
y_n^{\text{ideal}} = y_n^{\text{ideal}}(\mathbf{p}_0) = \sigma_0 \cdot \gamma(|\mathbf{p}_n-\mathbf{p}_0|) e^{j\Phi(|\mathbf{p}_n-\mathbf{p}_0|)},
\label{eq:reflector_ideal}
\end{equation}
where $\sigma_0$ is the constant radar cross-section of the corner reflector. Correspondingly, the actual received signal from the reflector includes antenna phase errors:
\begin{equation}
y_n^{\text{actual}} = y_n^{\text{actual}}(\mathbf{p}_0) = \sigma_0 \cdot \gamma(|\mathbf{p}_n-\mathbf{p}_0|) e^{j(\Phi(|\mathbf{p}_n-\mathbf{p}_0|) + \Delta\phi_n)}.
\label{eq:reflector_actual}
\end{equation}

For calibration, corner reflectors are typically placed at distances greater than 5m and directly facing the radar to satisfy far-field assumptions. In this case, propagation distances across the antenna array are approximately equal:
\begin{equation}
|\mathbf{p}_1-\mathbf{p}_0| \approx |\mathbf{p}_2-\mathbf{p}_0| \approx \cdots \approx |\mathbf{p}_N-\mathbf{p}_0|
\label{eq:equal_distances}
\end{equation}
Under these conditions, attenuation and phase delay terms are nearly constant across antennas. Therefore, the phase error of the $n$-th antenna can be directly estimated as:
\begin{equation}
\Delta\phi_n = \angle\left(y_n^{\text{actual}}(\mathbf{p}_0)\right) - \angle\left(y_{1}^{\text{actual}}(\mathbf{p}_0)\right)
\label{eq:phase_estimation}
\end{equation}
Corner reflectors can also be placed at an angle relative to the radar (see Supplementary Note III). Additionally, other calibration objects with well-characterized radio properties (such as metal plates~\cite{sprengUWBNearfieldMIMO2013}, spheres~\cite{guoMillimeterWave3DImaging2020,bleh100GHzFMCW2016}, and metal tea cans~\cite{liuMIMORadarCalibration2018}) can be used. However, all relevant scattering parameters ($\Omega$, $\sigma(\mathbf{p}_\omega; \mathbf{p}_n)$) should be known as priori, thus also requiring professional operation and measurement.

\begin{figure}[!htbp]
    \centering
    \includegraphics[width=0.9\linewidth]{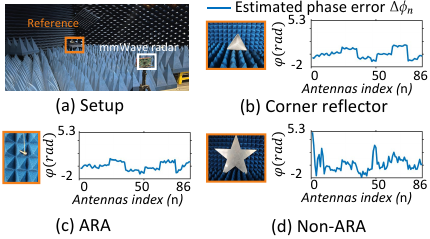}
    \caption{Phase error calibration comparison using different reference objects.}
    \label{fig3:calibration_comparison}
\end{figure}

\noindent \textbf{Results and limitations.} 
In October 2024, using a corner reflector 6 meters away, we measured phase error of a TI MMWCAS-RF-EVM radar that had not been calibrated for over a year, as shown in Fig.~\ref{fig3:calibration_comparison}(a) and (b). These measurements, based on the reliable corner reflector, are considered the groud truth of phase errors. 
However, such professional calibration procedures requiring corner reflector placement and maintenance are inconvenient and impractical for ordinary users like vehicle owners.


\subsubsection{Calibration Using Ambient Reflectors}

Several approaches attempt to leverage naturally occurring reflectors in the environment for radar calibration, avoiding the need for artificial reference objects. These methods fall into two main categories: \textit{Stable Scatterer Methods}~\cite{zhangRFSearchSearchingUnconscious2023,zhangMobi2Sense2022} select peaks with high signal strength in the radar range profile, and \textit{Permanent Scatterer Methods}~\cite{tagliaferriCooperativeCoherentMultistatic2024} identify reflectors maintaining consistent phase responses across multiple measurements. Both approaches estimate phase error using:

\begin{equation}
    \Delta\phi_n = \angle\left(y_n^{\text{actual}}(\mathbf{p}_a)\right) - \angle\left(y_{1}^{\text{actual}}(\mathbf{p}_a)\right) - \delta\phi_{ideal}(\mathbf{p}_a)
    \label{eq:env_phase_estimation}
\end{equation}

The critical limitation shared by both methods is their inability to accurately determine $\delta\phi_{ideal}(\mathbf{p}_a)$—the expected phase difference between adjacent antennas. Unlike corner reflectors with stable phase centers (Eq.~\eqref{eq:corner_reflector_condition}), environmental objects exhibit complex, unknown electromagnetic scattering properties that vary with slight changes in viewing angle. This fundamental electromagnetic constraint prevents these methods from achieving high-precision calibration required for mmWave systems, as they lack a rigorous framework to identify reflectors that truly satisfy Eq.~\eqref{eq:corner_reflector_condition}.

\noindent \textbf{Results and limitations.} Concurrently in October 2024, we performed calibration using strong environmental reflectors under identical conditions. Fig.~\ref{fig3:calibration_comparison}(d) reveals phase error measurements that diverge significantly from the ground truth (Fig.~\ref{fig3:calibration_comparison}(b)). This divergence shows that such methods, although convenient, lack the precision.

\subsection{Opportunity: Ambient Radio Anchors }
Our analysis of existing calibration approaches reveals a fundamental question: Can we develop a calibration method that combines the convenience of using environmental objects with the precision of corner reflectors? To answer this, we should first understand what makes corner reflectors effective for calibration.

\subsubsection{The Mathematical Basis of Effective Calibrators}

\begin{figure}[!htbp]
    \centering
    \includegraphics[width=0.9\linewidth]{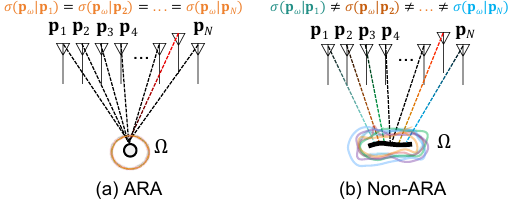}
    \caption{Defining Ambient Radio Anchors (ARAs).}
    \label{fig:define_ara}
\end{figure}

The effectiveness of corner reflectors stems from their consistent RCS across each receiving antenna. As illustrated in Fig.~\ref{fig:define_ara}, this property can be expressed mathematically as:
$\frac{\partial \sigma(\mathbf{p}_\omega; \mathbf{p}_n)}{\partial \mathbf{p}_n} = 0, \forall \omega \in \Omega$ (Eq.~\eqref{eq:corner_reflector_condition}).
We define ambient objects satisfying this condition as ARAs.

\subsubsection{Two Ways to Becoming an ARA}

Our key insight is that the ARA condition can be satisfied through two distinct physical mechanisms:

\noindent \textbf{(1) Through geometric structure:} Specific geometric configurations can create the phase stability required for ARAs. Corner reflectors exemplify this mechanism through their triple-orthogonal surfaces that reflect waves back toward the source regardless of incidence angle, creating a stable virtual phase center.

\noindent \textbf{(2) Through RCS size:} According to electromagnetic scattering theory, when an object's dimensions ($d$) approach the wavelength ($\lambda$, approximately 4mm at 77GHz), its scattering behavior transitions from the optical region to the resonance region. In this regime, the object's scattering pattern becomes dominated by resonance effects rather than geometric optics. This physical phenomenon creates a stable phase center because the electromagnetic field interactions concentrate at the object's center, producing consistent phase relationships across different viewing angles (Eq.~\eqref{eq:corner_reflector_condition}). Consequently, small metallic objects like screws, rivets, and electronic components naturally satisfy the ARA condition without specialized geometric design.

\subsubsection{Experimental Verification of Natural ARAs}

\begin{figure}[!htbp]
    \centering
    \includegraphics[width=0.93\linewidth]{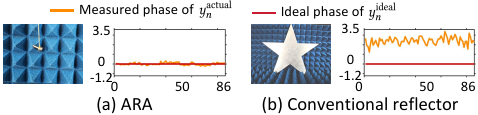}
    \caption{Phase response: ARA vs. Non-ARA.}
    \label{fig:ara_verification}
\end{figure}

We first verified that small RCS objects indeed qualify as ARAs, while traditional reference with strong reflections do not. Fig.~\ref{fig:ara_verification}(a) demonstrates that when testing with a pre-calibrated radar, a metallic object (screw) at 5m produces nearly identical phase responses across all antennas, satisfying the ARA definition.
In contrast, Fig.~\ref{fig:ara_verification}(b) shows that conventional reference targets (metal plate) generate inconsistent phase responses across different antennas.

Next, we validated that natural ARAs can calibrate radar systems as effectively as corner reflector. Specifically, we used the small metallic screw positioned 5 meters from the same radar in October 2024. Fig.~\ref{fig3:calibration_comparison}(c) reveals that this ARA produced phase error measurements remarkably consistent with the corner reflector ground truth (Fig.~\ref{fig3:calibration_comparison}(b)).

\begin{figure}[!htbp]
    \centering
    \includegraphics[width=0.95\linewidth]{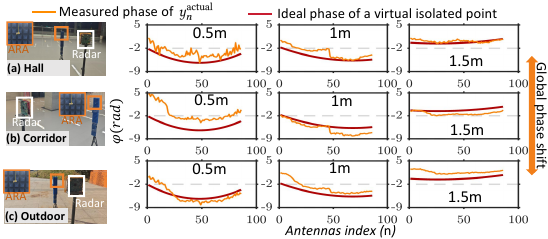}
    \caption{Experimental validation of VIP and PI. Each row validates VIP by showing ARAs exhibit phase characteristics similar to ideal point targets. Each column validates PI by demonstrating consistent ARA signatures across different environments with only global phase shifts.}
    \label{fig:vip_pi_validation}
\end{figure}

\begin{figure*}[!htbp]
    \centering
    \includegraphics[width=0.95\linewidth]{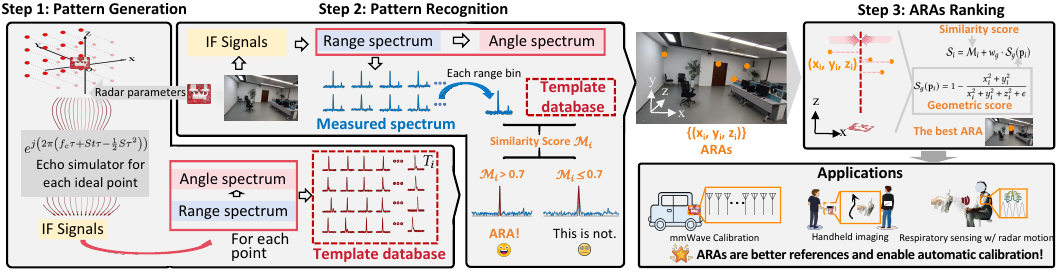}
    \caption{System overview. AutoCalib first generates theoretical templates, then matches these templates against measured radar data to recognize ARAs, and finally ranks candidates to select the optimal ARA for applications.}
    \label{fig:system_overview}
\end{figure*}

\subsubsection{Feasibility of Finding ARAs}

The above analysis has confirmed that ARAs are effective and exist, but their reliable identification in practical environments requires rigorous theoretical analysis. We must overcome two theoretical obstacles: (1) distinguishing objects with ARA properties from the multitude of reflectors with diverse electromagnetic characteristics, and (2) maintaining detection consistency across different environments.

Our analysis yields two critical properties of ARA that can solve the above obstacles and enable ARA detection:

\begin{lemma}[\textbf{Virtual Isolated Point (VIP)}]
    ARAs have radar response electromagnetically equivalent to that of an ideal isolated point (Proof in Supplementary Note IV).
\end{lemma}

\begin{lemma}[\textbf{Pattern Invariance (PI)}]
    ARAs maintain characteristic structure under different environments, experiencing only global phase shifts without structural distortion (Proof in Supplementary Note IV).
\end{lemma}

\noindent \textit{Experimental Validation.}
Each row in Fig.~\ref{fig:vip_pi_validation} validates the VIP property, demonstrating that ARAs exhibit phase characteristics quite similar to those of ideal point targets. Meanwhile, each column confirms the PI property, showing that different environments introduce only global phase variations without altering the relative phase relationships across antennas.

\section{AutoCalib Design}

VIP and PI provide a theoretical foundation for the practical detection of ARAs. According to VIP, ARAs generate antenna phase responses that precisely match those of ideal isolated point, enabling accurate mathematical modeling. Simultaneously, PI ensures that distinctive signatures of ARAs remain structurally consistent across varying environments, with only global phase shifts occurring even when surrounding electromagnetic conditions change substantially.

Together, these properties suggest a natural but powerful detection approach: by generating theoretical templates of ideal ARA signatures (leveraging VIP) and matching real-world radar measurements against these templates, we can reliably identify natural ARAs in everyday environments regardless of environmental conditions (thanks to PI).

\subsection{System Overview}

This idea leads us to AutoCalib, a template-matching framework for discovering and leveraging ARAs in radar calibration. However, there are three key challenges: (1) How to derive ideal templates that accurately represent theoretical ARA characteristics; (2) How to efficiently identify matching patterns within complex, noisy radar data; and (3) How to select optimal ARAs when multiple ARAs exist. To address these challenges, AutoCalib implements a systematic three-stage process (Fig.~\ref{fig:system_overview}):

\noindent \textbf{Pattern generation} first translates ARA theory into a searchable library of electromagnetic signatures based on the VIP property, creating precise template database of  how ideal point scatterers appear in radar measurements (Sec.~\ref{sec:step1}).

\noindent \textbf{Pattern recognition} transforms raw radar measurements into spatial spectrum representations and compare them against our theoretical templates to identify ARAs through their distinctive electromagnetic signatures, even when their reflections are weak relative to surrounding objects (Sec.~\ref{sec:step2}).

\noindent \textbf{ARAs ranking} refines candidates through a comprehensive scoring model that combines pattern matching quality with geometric considerations, prioritizing ARAs positioned to deliver maximum calibration precision (Sec.~\ref{sec:step3}).

\subsection{Pattern Generation} \label{sec:step1}

\subsubsection{Which Pattern? Spatial Spectrum}

To identify ARAs effectively, we must first determine which electromagnetic pattern representation to use. The pattern should be uniquely derivable from radar echos through invertible transformations. In radar signal processing, Fourier transforms are commonly applied to extract various spectral representations: range spectrum, Doppler spectrum, and spatial spectrum.

We selected spatial spectrum based on three key considerations:
\noindent \textbf{(1) Capturing antenna phase errors.} Spatial spectrum directly represents the critical inter-antenna phase relationships central to our calibration task, explicitly capturing the angular coherence properties that define ARAs.
\noindent \textbf{(2) Preserving ARA properties.} Only spatial spectrum fully satisfies both VIP and PI properties. Unlike range or Doppler spectra that change with environmental variations, spatial spectrum remains stable by effectively filtering interference from directions outside the ARA's position.
\noindent \textbf{(3) Balancing information and complexity.} While multi-dimensional spectral representations (e.g., 2D spatial-range spectrum) could provide additional information, 1D spatial spectrum offers a trade-off between discriminative capability and computational efficiency.

\begin{figure}[!htbp]
    \centering
    \includegraphics[width=0.9\linewidth]{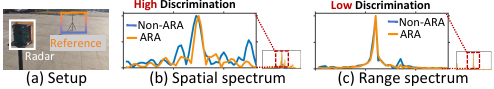}
    \caption{Spatial spectrum is a suitable template because it effectively discriminates ARA and non-ARA.}
    \label{fig:spatial_pattern_validation}
\end{figure}

To verify spatial spectrum's discriminative capability, we conducted controlled experiments comparing an ARA and a non-ARA object placed at identical angles and distances, as shown in Fig.~\ref{fig:spatial_pattern_validation}. Their range spectra exhibited 87\% similarity, making differentiation challenging. In contrast, their spatial spectra showed only 42\% similarity—a significant differential that enables robust classification, confirming spatial spectrum as a good pattern representation for ARA identification.

\subsubsection{Theoretical Pattern Generation}
To enable robust ARA identification, we develop a template database predicting how ideal ARAs at different positions would manifest in the spatial spectrum representation.

For an ideal ARA at position $\mathbf{p}_i$, according to VIP and PI, its theoretical response $y_n^{\text{ideal}}(\mathbf{p}_i)$ can be derived from well-studied radar signal model (detailed derivation in Supplementary Note V). We then transform this signal into a spatial spectrum template through our Fourier analysis framework:

\begin{equation}
    T_i = \mathcal{F}\{y_n^{\text{ideal}}(\mathbf{p}_i)\} = \frac{\mathcal{F}_a\{\mathcal{F}_r\{y_n^{\text{ideal}}(\mathbf{p}_i)\}\}}{\|\mathcal{F}_a\{\mathcal{F}_r\{y_n^{\text{ideal}}(\mathbf{p}_i)\}\}\|}
\end{equation}
where $\mathcal{F}$ represents our complete 2D Fourier transform process, with $\mathcal{F}_r$ denoting the range-domain transform and $\mathcal{F}_a$ the angular-domain transform across antenna elements.

\begin{algorithm}
    \footnotesize  
    \setlength{\abovedisplayskip}{3pt}  
    \setlength{\belowdisplayskip}{3pt}  
    \setlength{\lineskip}{3pt}  
    \caption{ARA Template Generation}
    \begin{algorithmic}[1]
    \Require Sensing volume $\mathcal{V}$, resolution $\delta$, radar configuration $\mathcal{R}$
    \Ensure Template database $\mathcal{T}$
    \State $\mathcal{T} \gets \emptyset$
    \For{each position $\mathbf{p}_i \in \mathcal{V}$ with step size $\delta$}
        \State Compute ideal radar response $y_n^{\text{ideal}}(\mathbf{p}_i)$ \Comment{Based on Eq. (28) in Supplementary Note V}
        \State Apply range-domain FFT: $S_r \gets \mathcal{F}_r\{y_n^{\text{ideal}}(\mathbf{p}_i)\}$
        \State Apply angular-domain FFT: $S_{r,a} \gets \mathcal{F}_a\{S_r\}$
        \State Extract spatial spectrum at target range: $S_{r_i,a} \gets S_{r,a}(r_i, :)$ 
        \State Normalize: $T_i \gets S_{r_i,a} / \|S_{r_i,a}\|$
        \State $\mathcal{T} \gets \mathcal{T} \cup \{(\mathbf{p}_i, T_i)\}$
    \EndFor
    \State \Return $\mathcal{T}$
    \end{algorithmic}
\end{algorithm}

\begin{figure}[!htbp]
    \centering
    \includegraphics[width=0.99\linewidth]{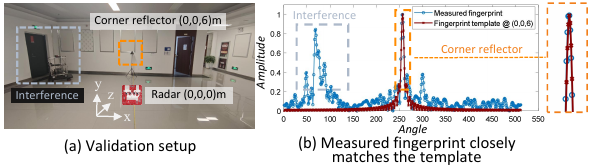}
    \caption{Validation of template match accuracy.}
    \label{fig:template_validation}
\end{figure}

To validate our template generation accuracy, we compared theoretical templates with measured ARA patterns from known corner reflectors. Fig.~\ref{fig:template_validation} shows the experimental setup (a) and the remarkable similarity between the measured spatial spectrum (blue) and theoretical template (red) for a corner reflector at position (0,0,6)m (b). This close correspondence confirms our model's accuracy even in complex multipath environments.

\subsection{Pattern Recognition} \label{sec:step2}

With the template database established, AutoCalib employs pattern recognition to detect potential ARAs from radar signal. For a radar with $N$ antennas measuring potential ARAs, we apply the same Fourier transform to process the received signals:

\begin{equation}
    S_{r,a} = \mathcal{F}\{y_n^{\text{actual}}\} = \mathcal{F}_a\{\mathcal{F}_r\{y_n^{\text{actual}}\}\}
\end{equation}
where we use the identical transformation $\mathcal{F}$ as in our template generation process.

\begin{algorithm}
    \footnotesize  
    \setlength{\abovedisplayskip}{3pt}  
    \setlength{\belowdisplayskip}{3pt}  
    \setlength{\lineskip}{3pt}  
    \caption{ARA Pattern Recognition}
    \begin{algorithmic}[1]
    \Require Measured signals $y_n^{\text{actual}}$, Template database $\mathcal{T}$, Threshold $\tau$
    \Ensure Set of detected ARAs $\mathcal{D}$
    \State $\mathcal{D} \gets \emptyset$
    \State Apply range-domain FFT: $S_r \gets \mathcal{F}_r\{y_n^{\text{actual}}\}$
    \State Apply angular-domain FFT: $S_{r,a} \gets \mathcal{F}_a\{S_r\}$
    \For{each range bin $r_i$}
        \State Extract spatial spectrum at range $r_i$: $S_{r_i,a} \gets S_{r,a}(r_i, :)$
        \State Normalize: $S_{\text{norm}} \gets S_{r_i,a} / \|S_{r_i,a}\|$
        \State $\mathcal{T}_{r_i} \gets$ \{Templates at distance $r_i$ from $\mathcal{T}$\}
        \For{each template $(\mathbf{p}_i, T_i) \in \mathcal{T}_{r_i}$}
            \State Compute similarity: $\mathcal{M}_{i} \gets \langle S_{\text{norm}}, T_i \rangle$
            \If{$\mathcal{M}_{i} \geq \tau$}
                \State $\mathcal{D} \gets \mathcal{D} \cup \{\mathbf{p}_i, \mathcal{M}_{i}\}$
            \EndIf
        \EndFor
    \EndFor
    \State \Return $\mathcal{D}$
    \end{algorithmic}
\end{algorithm}

The core of our detection approach is distance-specific pattern matching, which enhances both computational efficiency and detection accuracy. After obtaining the complete 2D range-angle spectrum $S_{r,a}$, we process each range bin independently by extracting its one-dimensional spatial spectrum and comparing it against templates from the same distance. The similarity metric is defined as:
\begin{equation}
    \mathcal{M}_i = \langle S_{\text{norm}}, T_i \rangle = \frac{\langle S_{r_i,a}, T_i \rangle}{\|S_{r_i,a}\|}
\end{equation}
where $\langle \cdot, \cdot \rangle$ denotes the inner product. 
This matching mechanism offers three key advantages: (1) it quantifies structural alignment between measured and theoretical patterns independent of signal strength; (2) by matching within specific range bins, it effectively isolates ARAs from interference caused by other reflectors, identifying ideal candidates for calibration applications; and (3) it enables parallel processing across range bins for computational efficiency.

A position $\mathbf{p}_i$ is classified as containing an ARA when:
\begin{equation}
    \mathcal{M}_i \geq \tau
\end{equation}
where $\tau$ is the detection threshold. We experimentally determined $\tau = 0.7$ as the threshold (Fig.~\ref{fig:ara_discovery2}).

Through \textit{Pattern Recognition}, AutoCalib can successfully identify weak but genuine ARAs while rejecting stronger non-ARA reflectors that lack the calibration characteristic defined in Eq.~\eqref{eq:corner_reflector_condition}.
\subsection{ARAs Ranking} \label{sec:step3}

After identifying potential ARAs through pattern recognition, AutoCalib should prioritize candidates based on their suitability for calibration.

\begin{algorithm}
    \footnotesize  
    \setlength{\abovedisplayskip}{3pt}  
    \setlength{\belowdisplayskip}{3pt}  
    \setlength{\lineskip}{3pt}  
    \caption{ARAs Ranking}
    \begin{algorithmic}[1]
    \Require Set of detected ARAs with matching scores $\mathcal{D} = \{(\mathbf{p}_i, \mathcal{M}_i)\}$
    \Ensure Ranked set of ARAs $\mathcal{A}$
    \State $\mathcal{A} \gets \emptyset$
    \For{each ARA position with score $(\mathbf{p}_i, \mathcal{M}_i)$}
        \State Compute geometric score: $\mathcal{S}_g(\mathbf{p}_i) \gets 1 - \frac{x_i^2 + y_i^2}{x_i^2 + y_i^2 + z_i^2 + \epsilon}$
        \State Compute final score: $\mathcal{S}_i \gets \mathcal{M}_i + w_g \cdot \mathcal{S}_g(\mathbf{p}_i)$
        \State $\mathcal{A} \gets \mathcal{A} \cup \{(\mathbf{p}_i, \mathcal{S}_i)\}$
    \EndFor
    \State Sort all $(\mathbf{p}_i, \mathcal{S}_i) \in \mathcal{A}$ by $\mathcal{S}_i$ in descending order
    \State \Return $\mathcal{A}$
    \end{algorithmic}
\end{algorithm}

\noindent \textbf{Geometric quality score.} 
While ARAs can theoretically serve as calibration references at various angles, those directly facing the radar provide superior performance~\cite{liHighresolutionHandheldMillimeterwave2024}. This preference stems from fundamental physical mechanisms (detailed in Supplementary Note VI): reduced radar cross-section (RCS) at oblique angles, and quadratically increasing phase compensation errors with angular deviation.

Therefore, we define a geometric quality score that captures these effects:
\begin{equation}
    \mathcal{S}_g(\mathbf{p}_i) = 1 - \frac{x_i^2 + y_i^2}{x_i^2 + y_i^2 + z_i^2 + \epsilon}
    \label{eq:geometric_score}
\end{equation}
where $(x_i, y_i, z_i)$ represents the ARA's position and $\epsilon$ is a small constant ($10^{-6}$) to prevent division by zero. This function reaches its maximum (1.0) when the ARA is directly in front of the radar and decreases as the ARA moves away from boresight. Notably, Eq.~\eqref{eq:geometric_score} also naturally excludes ARAs too close to the radar, ensuring selected candidates satisfy far-field conditions (distance $\geq$ 3.7m).

\noindent \textbf{Ranking mechanism.} 
The final comprehensive score combines pattern matching quality and geometric optimality:

\begin{equation}
    \mathcal{S}_i = \mathcal{M}_i + w_g \cdot \mathcal{S}_g(\mathbf{p}_i)
\end{equation}
where $w_g = 0.6$ balances pattern matching quality and geometric positioning, determined through microbenchmark experiments in Sec.~\ref{sec:534ImapctOfWeight}.

By integrating pattern recognition confidence with geometric optimality, AutoCalib effectively prioritizes ARAs that deliver highest calibration performance across diverse sensing applications.

\section{Evaluation}

This section first evaluates AutoCalib's ability to localize ARAs, then assesses its performance in mmWave radar calibration using the discovered ARAs. Following that, we evaluate the robustness of AutoCalib through micro-benchmarks, and finally explore the potential of applying AutoCalib to other phase calibration tasks.

\subsection{ARAs Discovery and Localization} \label{sec:evaluation1}

\subsubsection{Experimental Setup}
We test AutoCalib across 11 diverse environments including outdoor parking lots, building corridors, offices, and open spaces. In 8 scenarios, corner reflectors are explicitly placed as ARAs, and in 3 scenarios, small ARAs (screws) are placed. Our hardware setup employs a TI cascaded MIMO mmWave radar with 86 virtual antennas, while an LS-128 LiDAR provides ARAs' ground truth position (Fig.~\ref{fig:ara_discovery}). Detailed hardware specifications are provided in Supplementary Note VII.

\subsubsection{Baselines and Metrics}
We compare AutoCalib against stable scatterer detection~\cite{zhangRFSearchSearchingUnconscious2023,zhangMobi2Sense2022} that selects references based on signal strength and stability, and (2) Permanent scatterer identification~\cite{tagliaferriCooperativeCoherentMultistatic2024} that leverages phase stability. 
We evaluate using localization error, i.e., the distance between detected reflector positions and LiDAR ground truth.

\begin{figure}[!htbp]
    \centering
    \includegraphics[width=0.95\linewidth]{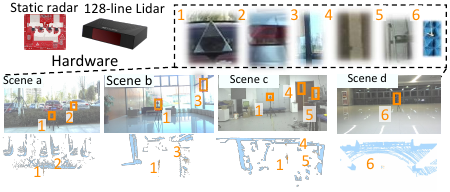}
    \caption{Qualitative results of ARA discovery. AutoCalib successfully identifies both placed references (\#1, \#6) and discovers natural ARAs (\#2-\#5) across four representative scenarios.}
    \label{fig:ara_discovery}
\end{figure}

\begin{figure}[!htbp]
    \begin{minipage}[t]{0.66\linewidth}
        \centering
        \includegraphics[width=\linewidth]{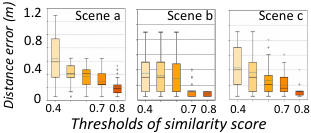}
        \caption{The impact of similarity threshold.}
        \label{fig:ara_discovery2}
    \end{minipage}
    \hfill
    \begin{minipage}[t]{0.32\linewidth}
        \centering
        \includegraphics[width=\linewidth]{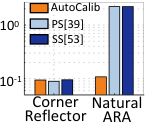}
        \caption{Quantitative results. (y-axis: error/m)}
        \label{fig:ara_discovery3}
    \end{minipage}
\end{figure}

\subsubsection{Qualitative Results}
Fig.~\ref{fig:ara_discovery} showcases AutoCalib's ability to identify ARA. In each scenarios, AutoCalib accurately identifies the corner reflector (\#1) and placed ARA (\#6) in the scene. Considering that AutoCalib is not a strength-based method, this shows that AutoCalib can match the reflection pattern of the ARAs.

Most importantly, we found several common objects (\#2, 3, 4, 5) that frequently serve as natural ARAs, including structural corners, metal fixtures, and small metallic objects like screws and rivets. Sec.~\ref{sec:case1} demonstrates that natural ARAs can be used in array calibration.

\subsubsection{The impact of Similarity Threshold}
Fig.~\ref{fig:ara_discovery2} demonstrates how similarity threshold balances detection precision against coverage. Thresholds below 0.5 yield high positioning errors (0.4-0.8m), while values above 0.7 consistently deliver sub-decimeter accuracy by effectively filtering non-ideal reflectors. Based on these results, we recommend 0.7 as the threshold for practical deployments.

\subsubsection{Quantitative Results}

Fig.~\ref{fig:ara_discovery3} reveals AutoCalib's exceptional capability to identify both strong and weak ARAs. With similarity threshold set to 0.7, AutoCalib achieves comparable performance to existing methods when localizing conventional corner reflectors (median error ~0.1m). However, in 3 environments with small ARA (metal screw), AutoCalib significantly outperforms baselines~\cite{zhangRFSearchSearchingUnconscious2023,zhangMobi2Sense2022,tagliaferriCooperativeCoherentMultistatic2024} by successfully detecting weaker ARAs that meet theoretical criteria but would be overlooked by existing approaches.

\subsection{mmWave Array Calibration} \label{sec:case1}

Next, we validate the performance of ARAs discovered by AutoCalib in mmWave radar calibration, which determines whether AutoCalib can effectively address the periodic recalibration problem identified in Section~\ref{sec:intro}.

\begin{figure}[!htbp]
\centering
    \includegraphics[width=0.95\linewidth]{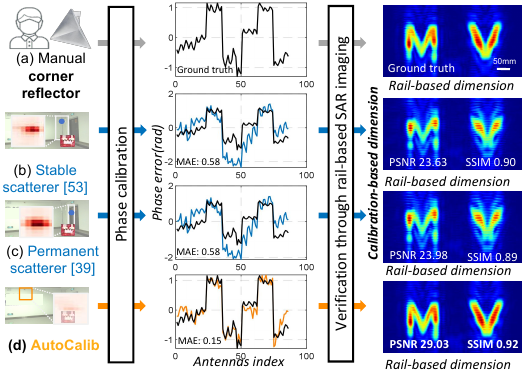}
    \caption{Qualitative results of radar calibration. } 
    \label{fig:calibration}
\end{figure}

\begin{table}[!htbp]
    \centering
    \caption{Comparison of radar array calibration (UC: Uncalibrated, PS: Permanent Scatter, SS: Stable Scatter, AC: AutoCalib, GT: Ground truth)}
    \label{tab:performance_array}
    \setlength{\tabcolsep}{3pt}
    \renewcommand{\arraystretch}{0.85}
    \small
    \begin{tabular*}{0.46\textwidth}{@{\extracolsep{\fill}}lccccc@{}}
    \toprule
     \textbf{Metrics} & \textbf{UC} & \textbf{PS\cite{tagliaferriCooperativeCoherentMultistatic2024}} & \textbf{SS\cite{zhangMobi2Sense2022}} & \textbf{AC} & \textbf{GT} \\
    \midrule
     MAE↓ & 0.61 & 0.57 & 0.56 & \textbf{0.17} & \underline{0.00} \\
     PSNR (dB)↑ & 19.06 & 23.80 & 24.70 & \textbf{28.15} & \underline{Inf.} \\
     SSIM↑ & 0.80 & 0.87 & 0.89 & \textbf{0.92} & \underline{1.00} \\
    \bottomrule
    \end{tabular*}
\end{table}

\subsubsection{Experiment Setup}
We compare AutoCalib against three baseline approaches: (1) Corner reflector-based calibration~\cite{xu3DHighResolutionImaging2024}, the current gold standard, (2) Stable~\cite{zhangMobi2Sense2022,zhangRFSearchSearchingUnconscious2023} and (3) Permanent scatterer~\cite{tagliaferriCooperativeCoherentMultistatic2024} methods. 

Our evaluation followed two steps: (1) Phase Error Estimation across 11 static scenes, measuring MAE = $\frac{1}{N}\sum_{n=1}^N |\Delta \phi_n - \Delta \phi_n^{\text{GT}}|$: the mean absolute difference between each method's estimated phase error vector $\Delta \phi_n$ and the corner reflector ground truth $\Delta \phi_n^{\text{GT}}$; (2) Further validation through MIMO-SAR imaging experiments, comparing PSNR and SSIM metrics against corner reflector results.

\subsubsection{Calibration Performance}
As shown in Fig.~\ref{fig:calibration}, AutoCalib achieves comparable performance to corner reflector calibration while significantly outperforming other baselines. Specifically, Fig.~\ref{fig:calibration}(a) shows that the mmWave radar has an average phase error of 0.64 rad per antenna. Using existing reference-based methods, whether stable scatterer or permanent scatterer, the estimated phase errors deviate from the true phase errors by 0.58 rad, indicating that calibration is almost ineffective. In contrast, the natural scatterer found by AutoCalib (a fire alarm in Fig.~\ref{fig:calibration}) successfully achieves a post-calibration MAE of only 0.17 rad. This represents a 74\% reduction compared to uncalibrated arrays, while outperforming existing ambient scatterer methods by 83\%.

\begin{figure}[!htbp]
    \centering
    \includegraphics[width=0.99\linewidth]{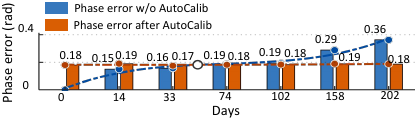}
    \caption{Minimum calibration interval assessment.}
    \label{fig:min_calibration_interval}
\end{figure}

\begin{figure}[!htbp]
    \centering
    \includegraphics[width=0.99\linewidth]{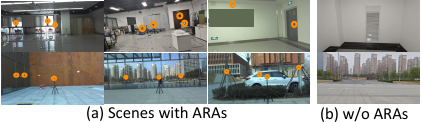}
    \caption{ARA presence across environments.}
    \label{fig11:araEnvironments}
\end{figure}

\begin{figure*}[!htbp]
    \centering
    \includegraphics[width=0.9\linewidth]{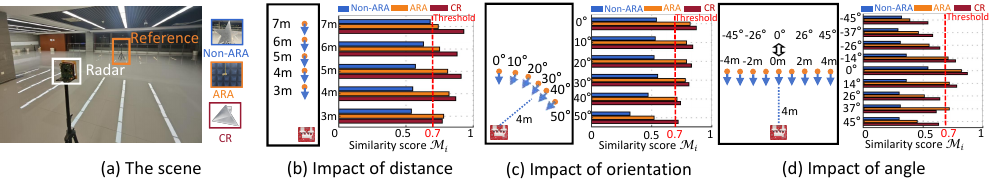}
    \caption{Evaluating AutoCalib under different distances, orientations and angles.}
    \label{fig:dynamic_impacts}
\end{figure*}

The rail-based MIMO-SAR imaging results in the third column in Fig.~\ref{fig:calibration} visually confirm this improvement. Across all test environments (Tab.~\ref{tab:performance_array}), AutoCalib achieves excellent image quality (PSNR: 28.15dB, SSIM: 0.92), closely matching corner reflector ground truth and significantly outperforming conventional approaches.

Notably, the most effective calibration references are not necessarily the strongest reflectors (Fig.~\ref{fig:calibration}). Unlike existing methods focused on signal strength, AutoCalib prioritizes geometric ideality. This fundamental difference explains its superior performance: \textit{weak but geometrically ideal scatterers provide more reliable phase references than strong but non-ideal reflectors.}

\subsubsection{Minimum Calibration Interval.}
We analyzed minimum effective calibration intervals for AutoCalib. Specifically, we conducted experiments with increasing time periods (Fig.~\ref{fig:min_calibration_interval}). We found that calibration intervals shorter than 74 days produced negligible improvements, while a 158-day interval successfully reduced phase errors by 34\%. 
This suggests that future large-array mmWave radars can incorporate AutoCalib-based routines to automatically perform calibration during idle periods, thereby promptly eliminating phase drift.

\subsection{Microbenchmark Evaluation}

Next, we evaluate AutoCalib's robustness under various influencing factors.

\subsubsection{ARA Presence Across Environments}

Understanding ARA availability across different environments is crucial for practical deployment. We evaluated ARA presence in diverse scenes, as shown in Fig.~\ref{fig11:araEnvironments}. Our results demonstrate that ARAs exist in most common environments (Fig.~\ref{fig11:araEnvironments}(a)), including hallways, offices, and spaces with furniture or structural elements. However, ARAs are scarce in certain specialized environments (Fig.~\ref{fig11:araEnvironments}(b)), such as completely empty rooms or open outdoor spaces with no objects. These results show that AutoCalib can satisfy calibration requirements in the vast majority of practical application scenarios.

\subsubsection{Impact of Different Distance}

We evaluated the effect of target distance on AutoCalib's performance. Specifically, we used three different reflectors (shown in Fig.~\ref{fig:dynamic_impacts}(a)): a metal plate, traditional reference object; a screw, deliberately placed small ARA; and a corner reflector, with a stable phase center. By positioning these reflectors at varying distances from the radar, we assessed distance-related impacts on AutoCalib. As shown in Fig.~\ref{fig:dynamic_impacts}(b), AutoCalib consistently identified both corner reflectors and ARAs effectively across the 3-7m range, demonstrating excellent distance robustness.

Interestingly, the metal plate's similarity score increased with distance, suggesting two important implications: (1) at greater distances, traditional strong scatterers like metal plates behave more like ARAs, which aligns with the superior performance of traditional methods in remote sensing~\cite{crosettoPersistentScattererInterferometry2016}; (2) AutoCalib works effectively in both near and far fields, while scatterers identified by traditional methods struggle to perform well in near-field applications.

\subsubsection{Impact of Different Orientations}
We then evaluated AutoCalib's performance when reflectors and radar were positioned at different orientations, with the reflectors still facing the radar, as shown in Fig.~\ref{fig:dynamic_impacts}(c). The results indicate that orientation has minimal impact on AutoCalib's performance until reaching approximately 40$^\circ$, at which point AutoCalib struggles to effectively detect the placed ARA. We attribute this limitation primarily to the radar's limited FOV, which results in insufficient SNR at 40$^\circ$. This excellent robustness mainly benefits from RCS stability: when reflectors face the radar, as long as the angle does not change significantly, the RCS remains stable from the radar's perspective, enabling consistent detection.

\begin{figure}[!htbp]
    \centering
    \includegraphics[width=0.92\linewidth]{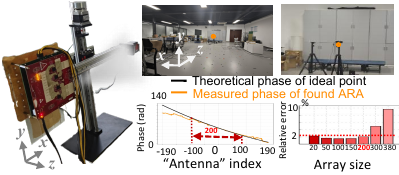}
    \caption{Impact of antenna array size.}
    \label{fig:antenna_array_size}
\end{figure}

\subsubsection{Impact of Viewing Angle}

We evaluated AutoCalib's performance when reflectors are not directly facing the radar, as shown in Fig.~\ref{fig:dynamic_impacts}(d). The reflector was positioned horizontally from $-4$m to $4$m at a distance of $4$m from the radar, creating viewing angle variations between $-45^\circ$ and $45^\circ$.
Fig.~\ref{fig:dynamic_impacts}(d) shows that both corner reflectors and ARAs maintain performance within a narrow angular range when using AutoCalib. Once the angle exceeds $14^\circ$, localization becomes difficult. This phenomenon occurs because angular changes significantly affect the reflector's RCS, severely altering reflection characteristics and destabilizing the phase center, thus failing to maintain the properties described in Eq.~\eqref{eq:corner_reflector_condition}.

Interestingly, when testing with metal plates (non-ARAs), AutoCalib successfully tracks them at certain oblique angles (37°), suggesting these objects exhibit ARA-like properties at specific orientations. This shows that AutoCalib identifies environmental features based on their radio characteristics rather than object type.

\subsubsection{Impact of Antenna Array Size} \label{sec:impactArray_size}

We then validate AutoCalib's applicability to large antenna arrays. Since physical arrays with 100+ antennas are uncommon, we used a precision rail system (specifications in Supplementary Note VII) moving at constant speed across 380 antenna positions to simulate a large linear array. After identifying ARAs with this simulated array, we compared the measured phase progression against theoretical point scatterer responses.

As shown in Fig.~\ref{fig:antenna_array_size}, the measured ARA phase response closely matches the theoretical ideal point response for approximately the first 200 antennas. This demonstrates AutoCalib's effectiveness for arrays with up to 200 elements, significantly exceeding the requirements of current commercial mmWave radar systems.

\subsubsection{Impact of Phase Error} \label{sec:impactPhase_error}

To assess AutoCalib's robustness against radar phase drift, we evaluated system performance under varying degrees of antenna phase error. As shown in Fig.~\ref{fig16:Maxphase_error}, the similarity score decreases as phase error increases, with performance degrading significantly when the average per-antenna phase error (MAE) exceeds 0.35 rad. This threshold suggests the maximum calibration interval for AutoCalib is approximately 700 days (calculated as 0.35 rad / 0.0005 rad/day), meaning that 4-chip radar systems that have not been calibrated for more than 700 days would not be suitable for AutoCalib and require calibration using professional corner reflector equipment.

\subsection{Case Studies} \label{sec:caseStudies}

The ability to correct phase distortions makes it valuable for various sensing applications affected by phase errors. Here, we validate its performance in two challenging applications: handheld imaging \cite{liIFNetDeepImaging2024,liHighresolutionHandheldMillimeterwave2024,alvarez-narciandiShorttimeCoherentIncoherent2024,schellbergMmSightRobustMillimeterWave2023,saadatMilliCamHandheldMillimeterWave2020} and respiratory sensing during radar motion~\cite{changMSenseBoostingWireless2024,zhangSinglePointMultiPointReflection2024,zhangMobi2Sense2022,chenCoSenseExploitingCooperative2024,zhangRFSearchSearchingUnconscious2023}, both of which suffer from phase errors introduced by radar movement.

\subsubsection{Case I: Handheld mmWave Imaging}

\begin{figure}[!htbp]
\centering
\includegraphics[width=0.92\linewidth]{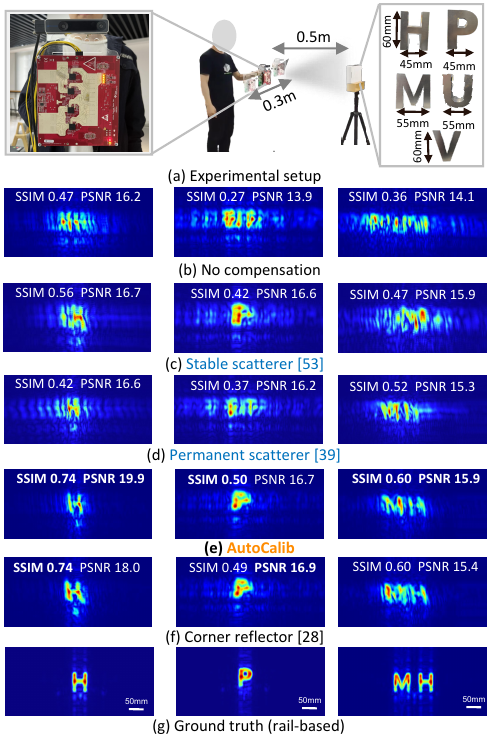}
\caption{Handheld mmWave imaging performance.}
\label{fig:handheld}
\end{figure}

Precise motion compensation is critical yet particularly challenging in handheld imaging due to the near-field nature.

\noindent \textbf{Experimental setup.} As shown in Fig.~\ref{fig:handheld}(a), our system employs a handheld TI cascaded mmWave radar to image letter patterns from 0.5m distance, with Intel RealSense T265 providing coarse position tracking. We collect data from 18 different objects. The detailed hardware specifications can be found in Supplementary Note VII.

\noindent \textbf{Performance evaluation.} We compare AutoCalib with uncalibrated, stable scatterer~\cite{zhangMobi2Sense2022,zhangRFSearchSearchingUnconscious2023}, permanent scatterer~\cite{tagliaferriCooperativeCoherentMultistatic2024}, and corner reflector (ground truth) approaches~\cite{liHighresolutionHandheldMillimeterwave2024}. Fig.~\ref{fig:handheld}(b) illustrates the comparative imaging results, while quantitative metrics are presented in Tab.~\ref{tab:performance_imaging_resp}. 

AutoCalib significantly outperforms uncalibrated processing (PSNR improvement of 2.44dB, SSIM improvement of 0.18) and conventional calibration methods, achieving near-identical quality to corner reflector calibration (99\% of ground truth PSNR and 97\% of ground truth SSIM). This is particularly impressive considering handheld imaging's sensitivity to phase calibration errors, as detailed in Supplementary Note VI.

\begin{figure}[!htbp]
    \begin{minipage}[t]{0.39\linewidth}
        \centering
        \includegraphics[width=\linewidth]{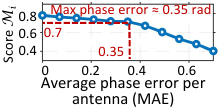}
        \caption{Impact of phase error.}
        \label{fig16:Maxphase_error}
    \end{minipage}
    \hfill
    \begin{minipage}[t]{0.58\linewidth}
        \centering
        \includegraphics[width=\linewidth]{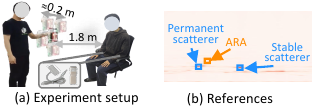}
        \caption{Respiratory sensing setup and references.}
        \label{fig:respiratory}
    \end{minipage}
\end{figure}

\subsubsection{Case II: Respiratory Sensing}

We also evaluate AutoCalib in contactless respiratory sensing with handheld mmWave radar, a challenging application due to subtle chest movements (2-5mm) and radar motion~\cite{zhangMobi2Sense2022,zhangRFSearchSearchingUnconscious2023}.

\noindent \textbf{Implementation.}
As shown in Fig.~\ref{fig:respiratory}(a), we evaluate respiratory sensing with a handheld radar monitoring a seated subject from 1.8m away, while the radar moves randomly within a 0.4m diameter range. Ground truth data is collected using a respiratory belt Supplementary Note VII. The dataset includes 15-minute recordings from 5 subjects.

\noindent \textbf{Performance evaluation.} Fig.~\ref{fig:respiratory}(b) visualizes the reference points identified by different methods, showing how AutoCalib successfully locates previously overlooked weak but effective ARAs. Tab.~\ref{tab:performance_imaging_resp} summarizes the quantitative results, where AutoCalib achieves a breathing rate error of only 0.41 breaths per minute (BPM), matching the performance of methods specifically designed for respiratory sensing tasks.

\begin{table}[!htbp]
    \centering
    \caption{Performance comparison}
    \label{tab:performance_imaging_resp}
    \setlength{\tabcolsep}{3pt}
    \renewcommand{\arraystretch}{0.85}
    \small
    \begin{tabular*}{0.46\textwidth}{@{\extracolsep{\fill}}p{1.2cm}lcccc c@{}}
    \toprule
     & \textbf{Metrics} & \textbf{UC} & \textbf{PS\cite{tagliaferriCooperativeCoherentMultistatic2024}} & \textbf{SS\cite{zhangMobi2Sense2022}} & \textbf{AC} & \textbf{GT} \\
    \midrule
    \multirow{2}{*}{\raggedright\textbf{Imaging}} 
     & PSNR (dB)↑ & 15.60 & 15.88 & 16.02 & \textbf{18.04} & \underline{18.24} \\
     & SSIM↑ & 0.45 & 0.54 & 0.51 & \textbf{0.63} & \underline{0.65} \\
    \midrule
    \textbf{Resp.} 
     & BPM Err.↓ & $\geq$10 & 0.42 & 0.41 & \textbf{0.41} & \underline{0.39} \\
    \bottomrule
    \end{tabular*}
\end{table}

These case studies demonstrate that AutoCalib's performance using discovered ARAs is comparable to corner reflector-based methods, validating its versatility across different applications requiring phase calibration.

\subsubsection{Impact of Weight $w_g$} \label{sec:534ImapctOfWeight}

\begin{figure}[!htbp]
    \centering
    \includegraphics[width=0.95\linewidth]{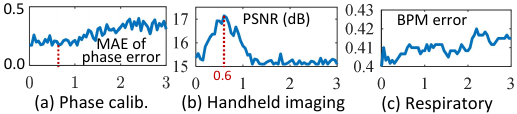}
    \caption{Impact of geometric weight $w_g$ on application performance.}
    \label{fig22:weight_impact}
\end{figure}

The weight parameter $w_g$ balances pattern matching and geometric positioning when selecting ARAs. Fig.~\ref{fig22:weight_impact} shows application-specific sensitivities to this parameter. Array calibration performs best with lower $w_g$ values, prioritizing pattern quality. Handheld imaging achieves optimal PSNR at $w_g=0.6$, requiring balanced consideration of both factors. Respiratory sensing shows minimal sensitivity to $w_g$ variations. Based on these results, we selected $w_g=0.6$ as our system's default value to support multiple applications effectively.

\subsection{Complexity Analysis}

ARA Template Generation is performed offline as preprocessing, leaving runtime complexity dependent only on Pattern Recognition and Ranking. With parallel implementation across range bins, this reduces to $O(|\mathcal{T}_{r_i}| + K \log K)$. Using a spatial resolution of 5cm×5cm×5cm on an Intel i7-12700K processor, AutoCalib completes calibration in 4.2 seconds, making it suitable for everyday applications. Detailed analysis is provided in Supplementary Note VIII.

\section{Limitations and Discussion}

\noindent \textbf{Applicable antenna array size.} In Sec.~\ref{sec:impactArray_size}, AutoCalib maintains reliable performance with arrays containing up to 200 antennas, making it suitable for most current and near-future commercial mmWave systems. For even larger arrays with 200+ elements, phase response degradation suggests the need for advanced techniques such as sub-array calibration or hierarchical approaches. We leave this as future work.

\noindent \textbf{Broader applications and frequency ranges.} While Sec.~\ref{sec:caseStudies} demonstrated AutoCalib across three case studies, its potential extends beyond mmWave radar applications. Future work could enhance rotational radar systems through improved phase calibration~\cite{laiEnablingVisualRecognition2024} and alignment~\cite{zhangMmRotationUnlockingVersatility2025}, enable higher spatial resolution physiological sensing~\cite{changMSenseBoostingWireless2024,zhangSinglePointMultiPointReflection2024}, strengthen passive sign-based mmWave navigation~\cite{iizukaMilliSignMmWaveBasedPassive2023}, and even calibrating mmWave communication system~\cite{nguyenDeepLearningFramework2023}. Moreover, the underlying principles of Virtual Isolated Points (VIP) and Scene Invariance (SI) are frequency-independent, suggesting potential applications across different wavelengths, from WiFi to THz, each presenting unique opportunities and challenges. ARAs hold promise across these diverse domains and frequency ranges by providing naturally occurring point-like references maintaining precise phase relationships.

\noindent \textbf{Multiple ARA utilization.} The current implementation of AutoCalib utilizes a single highest-scoring ARA for calibration. While this approach provides sufficient accuracy for most applications, environments with multiple ARAs offer opportunities for further performance enhancement. Future work could explore weighted calibration schemes that leverage multiple ARAs simultaneously.

\section{Conclusion}

This paper first introduced Ambient Radio Anchors (ARAs), naturally occurring scatterers providing stable phase references despite weak reflection strength. Then, based on ARAs' property, we presented AutoCalib, the first method capable of identifying these ideal phase references, overcoming existing approaches' limitation of focusing solely on strong reflectors while missing geometrically superior calibration sources. In radar applications, AutoCalib achieved accuracy comparable to corner reflector methods while significantly outperforming other reflector-free techniques. Experiments in handheld imaging demonstrated AutoCalib's effectiveness in enabling reflector-free mmWave sensing with professional-grade precision, facilitating practical deployment of high-resolution mmWave sensing in everyday environments.

\bibliographystyle{IEEEtran}
\bibliography{ruixu_new}

\begin{thebibliography}{10}
\providecommand{\url}[1]{#1}
\csname url@samestyle\endcsname
\providecommand{\newblock}{\relax}
\providecommand{\bibinfo}[2]{#2}
\providecommand{\BIBentrySTDinterwordspacing}{\spaceskip=0pt\relax}
\providecommand{\BIBentryALTinterwordstretchfactor}{4}
\providecommand{\BIBentryALTinterwordspacing}{\spaceskip=\fontdimen2\font plus
\BIBentryALTinterwordstretchfactor\fontdimen3\font minus \fontdimen4\font\relax}
\providecommand{\BIBforeignlanguage}[2]{{%
\expandafter\ifx\csname l@#1\endcsname\relax
\typeout{** WARNING: IEEEtran.bst: No hyphenation pattern has been}%
\typeout{** loaded for the language `#1'. Using the pattern for}%
\typeout{** the default language instead.}%
\else
\language=\csname l@#1\endcsname
\fi
#2}}
\providecommand{\BIBdecl}{\relax}
\BIBdecl

\bibitem{delafrooznorooziZeroShotAccurateMmWave2024}
O.~Delafrooz~Noroozi, H.~Guo, R.~Shen, Z.~Shao, H.~Chen, K.~Sengupta, Y.~Ghasempour, and U.~Madhow, ``Zero-{{Shot Accurate mmWave Antenna Array Calibration}} in the {{Wild}},'' in \emph{Proceedings of the 30th {{Annual International Conference}} on {{Mobile Computing}} and {{Networking}}}, ser. {{ACM MobiCom}} '24, Dec. 2024, pp. 1938--1945.

\bibitem{tagliaferriCooperativeCoherentMultistatic2024}
D.~Tagliaferri, M.~Manzoni, M.~Mizmizi, S.~Tebaldini, A.~V. {Monti-Guarnieri}, C.~M. Prati, and U.~Spagnolini, ``Cooperative {{Coherent Multistatic Imaging}} and {{Phase Synchronization}} in {{Networked Sensing}},'' \emph{IEEE Journal on Selected Areas in Communications}, pp. 1--1, 2024.

\bibitem{yaoRadarPerceptionAutonomous2023}
S.~Yao, R.~Guan, Z.~Peng, C.~Xu, Y.~Shi, Y.~Yue, E.~G. Lim, H.~Seo, K.~L. Man, X.~Zhu \emph{et~al.}, ``Radar {{Perception}} in {{Autonomous Driving}}: {{Exploring Different Data Representations}},'' \emph{arXiv preprint arXiv:2312.04861}, pp. 1--22, 2023.

\bibitem{huangOverviewSignalProcessing2023}
Y.~Huang, H.~Zhang, L.~Lan, K.~Deng, Y.~Yang, R.~Zhang, J.~Liu, Y.~Zhang, Y.~Wang, R.~Zhou, J.~Xu, X.~Xi, X.~Zhang, K.~Zheng, Y.~Liu, and W.~Hong, ``Overview of signal processing techniques for automotive millimeter-wave radar,'' \emph{Journal of Radars}, vol.~12, no.~5, pp. 923--970, 2023.

\bibitem{zhuMaliciousAttacksMultiSensor2024}
Y.~Zhu, C.~Miao, H.~Xue, Y.~Yu, L.~Su, and C.~Qiao, ``Malicious {{Attacks}} against {{Multi-Sensor Fusion}} in {{Autonomous Driving}},'' in \emph{Proceedings of the {{Annual International Conference}} on {{Mobile Computing And Networking}} ({{MobiCom}})}, May 2024, pp. 436--451.

\bibitem{zhangSurveyMmWaveBasedHuman2023}
J.~Zhang, R.~Xi, Y.~He, Y.~Sun, X.~Guo, W.~Wang, X.~Na, Y.~Liu, Z.~Shi, and T.~Gu, ``A {{Survey}} of {{mmWave-Based Human Sensing}}: {{Technology}}, {{Platforms}} and {{Applications}},'' \emph{IEEE Communications Surveys \& Tutorials}, vol.~25, no.~4, pp. 2052--2087, 2023.

\bibitem{zhangMonitoringLongtermCardiac2024}
B.-B. Zhang, D.~Zhang, Y.~Li, Z.~Lu, J.~Chen, H.~Wang, F.~Zhou, Y.~Pu, Y.~Hu, L.-K. Ma, Q.~Sun, and Y.~Chen, ``Monitoring long-term cardiac activity with contactless radio frequency signals,'' \emph{Nature Communications}, vol.~15, no.~1, p. 10598, Dec. 2024.

\bibitem{shastriReviewMillimeterWave2022}
A.~Shastri, N.~Valecha, E.~Bashirov, H.~Tataria, M.~Lentmaier, F.~Tufvesson, M.~Rossi, and P.~Casari, ``A {{Review}} of {{Millimeter Wave Device-Based Localization}} and {{Device-Free Sensing Technologies}} and {{Applications}},'' \emph{IEEE Communications Surveys \& Tutorials}, vol.~24, no.~3, pp. 1708--1749, 2022.

\bibitem{huContactlessArterialBlood2024}
Q.~Hu, Q.~Zhang, H.~Lu, S.~Wu, Y.~Zhou, Q.~Huang, H.~Chen, Y.-C. Chen, and N.~Zhao, ``Contactless {{Arterial Blood Pressure Waveform Monitoring}} with {{mmWave Radar}},'' \emph{Proceedings of the ACM on Interactive, Mobile, Wearable and Ubiquitous Technologies (IMWUT)}, vol.~8, no.~4, pp. 178:1--178:29, Nov. 2024.

\bibitem{wangPrecise2D3D2023}
Y.~Wang, J.~Su, T.~Fukuda, M.~Tonouchi, and H.~Murakami, ``Precise {{2D}} and {{3D Fluoroscopic Imaging}} by {{Using}} an {{FMCW Millimeter-Wave Radar}},'' \emph{IEEE Access}, vol.~11, pp. 84\,027--84\,034, 2023.

\bibitem{albabaSidelobesGhostTargets2024}
A.~Albaba, M.~Bauduin, A.~Sakhnini, H.~Sahli, and A.~Bourdoux, ``Sidelobes and {{Ghost Targets Mitigation Technique}} for {{High-Resolution Forward-Looking MIMO-SAR}},'' \emph{IEEE Transactions on Radar Systems}, vol.~2, pp. 237--250, 2024.

\bibitem{liHighresolutionHandheldMillimeterwave2024}
Y.~Li, D.~Zhang, R.~Geng, Z.~Lu, Z.~Wu, Y.~Hu, Q.~Sun, and Y.~Chen, ``A high-resolution handheld millimeter-wave imaging system with phase error estimation and compensation,'' \emph{Communications Engineering}, vol.~3, no.~1, pp. 1--11, 2024.

\bibitem{instrumentsAWR1843AutomotiveRadar2018}
T.~Instruments, ``{{AWR1843}} automotive radar sensor data sheet,'' Texas Instruments, Tech. Rep., 2018.

\bibitem{wangRFGymCareIntroducingRespiratory2024}
J.~Wang, D.~Zhang, B.~Zhang, J.~Chen, Y.~Hu, and Y.~Chen, ``{{RF-GymCare}}: {{Introducing Respiratory Prior}} for {{RF Sensing}} in {{Gym Environments}},'' \emph{Proc. ACM Interact. Mob. Wearable Ubiquitous Technol.}, vol.~8, no.~3, pp. 135:1--135:28, Sep. 2024.

\bibitem{bauderMMMUREMmWaveBasedMultiSubject2024}
C.~Bauder, A.-K. Moadi, V.~Rajagopal, T.~Wu, J.~Liu, and A.~E. Fathy, ``{{MM-MURE}}: {{mmWave-Based Multi-Subject Respiration Monitoring}} via {{End-to-End Deep Learning}},'' \emph{IEEE Journal of Electromagnetics, RF and Microwaves in Medicine and Biology}, pp. 1--13, 2024.

\bibitem{caoMmCLIPBoostingMmWavebased2024}
Q.~Cao, H.~Xue, T.~Liu, X.~Wang, H.~Wang, X.~Zhang, and L.~Su, ``{{mmCLIP}}: {{Boosting mmWave-based Zero-shot HAR}} via {{Signal-Text Alignment}},'' in \emph{Proceedings of the 22nd {{ACM Conference}} on {{Embedded Networked Sensor Systems}}}, ser. {{SenSys}} '24, Nov. 2024, pp. 184--197.

\bibitem{wengLargeModelSmall2024}
Y.~Weng, G.~Wu, T.~Zheng, Y.~Yang, and J.~Luo, ``Large {{Model}} for {{Small Data}}: {{Foundation Model}} for {{Cross-Modal RF Human Activity Recognition}},'' in \emph{Proceedings of the {{ACM Conference}} on {{Embedded Networked Sensor Systems}} ({{SenSys}})}, Oct. 2024, pp. 1--14.

\bibitem{chengNovelRadarPoint2022}
Y.~Cheng, J.~Su, M.~Jiang, and Y.~Liu, ``A {{Novel Radar Point Cloud Generation Method}} for {{Robot Environment Perception}},'' \emph{IEEE Transactions on Robotics}, vol.~38, no.~6, pp. 1--20, 2022.

\bibitem{gengDREAMPCDDeepReconstruction2024}
R.~Geng, Y.~Li, D.~Zhang, J.~Wu, Y.~Gao, Y.~Hu, and Y.~Chen, ``{{DREAM-PCD}}: {{Deep Reconstruction}} and {{Enhancement}} of {{mmWave Radar Pointcloud}},'' \emph{IEEE Transactions on Image Processing}, vol.~33, pp. 6774--6789, 2024.

\bibitem{caiMilliPCDTraditionalVision2023}
{\relax PINGPING}.~CAI and {\relax SANJIB}.~SUR, ``{{MilliPCD}}: {{Beyond Traditional Vision Indoor Point Cloud Generation}} via {{Handheld Millimeter-Wave Devices}},'' in \emph{Proceedings of the {{ACM}} on {{Interactive}}, {{Mobile}}, {{Wearable}} and {{Ubiquitous Technologies}} ({{IMWUT}})}, vol.~6, 2023, pp. 1--24.

\bibitem{karimian-sichaniAntennaArrayWaveform2024}
N.~{Karimian-Sichani}, M.~{Alaee-Kerahroodi}, B.~S. Mysore Rama~Rao, E.~Mehrshahi, and S.~A. Ghorashi, ``Antenna {{Array}} and {{Waveform Design}} for 4-{{D-Imaging mmWave MIMO Radar Sensors}},'' \emph{IEEE Transactions on Aerospace and Electronic Systems}, vol.~60, no.~2, pp. 1848--1864, Apr. 2024.

\bibitem{turkmen223276GHzCascadable2023}
E.~Turkmen, I.~K. Aksoyak, P.~Djondovic, S.~B. Yilmaz, W.~Debski, W.~Winkler, and A.~{\c C}. Ulusoy, ``A 223--276-{{GHz Cascadable FMCW Transceiver}} in 130-nm {{SiGe BiCMOS}} for {{Scalable MIMO Radar Arrays}},'' \emph{IEEE Transactions on Microwave Theory and Techniques}, vol.~71, no.~12, pp. 5393--5412, Dec. 2023.

\bibitem{zhengEnhancingMmWaveRadar2024}
K.~Zheng, W.~Zhao, T.~Woodford, R.~Zhao, X.~Zhang, and Y.~Hua, ``Enhancing {{mmWave Radar Sensing Using}} a {{Phased-MIMO Architecture}},'' in \emph{Proceedings of the {{International Conference}} on {{Mobile Systems}}, {{Applications}}, and {{Services}} ({{MobiSys}})}, ser. {{MOBISYS}} '24, 2024, pp. 56--69.

\bibitem{tsaregorodtsevAutomatedAutomotiveRadar2023}
A.~Tsaregorodtsev, M.~Buchholz, and V.~Belagiannis, ``Automated {{Automotive Radar Calibration}} with {{Intelligent Vehicles}},'' in \emph{Proceedings of the {{European Signal Processing Conference}} ({{EUSIPCO}})}, Sep. 2023, pp. 800--804.

\bibitem{heiningOvertheAirCalibrationMmW2023}
S.~Heining, R.~Feger, C.~Wagner, and A.~Stelzer, ``Over-the-{{Air Calibration}} of {{mmW Imaging Radars Using Uncorrelated Continuous Wave Signals}},'' \emph{IEEE Journal of Microwaves}, vol.~3, no.~3, pp. 970--979, Jul. 2023.

\bibitem{xu3DHighResolutionImaging2024}
G.~Xu, Y.~Chen, A.~Ji, B.~Zhang, C.~Yu, and W.~Hong, ``3-{{D High-Resolution Imaging}} and {{Array Calibration}} of {{Ground-Based Millimeter-Wave MIMO Radar}},'' \emph{IEEE Transactions on Microwave Theory and Techniques}, pp. 4919--4931, 2024.

\bibitem{raphaeliChallengesAutomotiveMIMO2023}
D.~Raphaeli and I.~Bilik, ``Challenges in {{Automotive MIMO Radar Calibration}} in {{Anechoic Chamber}},'' \emph{IEEE Transactions on Aerospace and Electronic Systems}, vol.~59, no.~5, pp. 6205--6214, Oct. 2023.

\bibitem{sprengUWBNearfieldMIMO2013}
T.~Spreng, U.~Prechtel, B.~Sch{\"o}nlinner, V.~Ziegler, A.~Meusling, and U.~Siart, ``{{UWB}} near-field {{MIMO}} radar: {{Calibration}}, measurements and image reconstruction,'' in \emph{2013 {{European Radar Conference}}}, Oct. 2013, pp. 33--36.

\bibitem{liuNearFieldCalibrationMillimeterWave2024}
J.~Liu, K.~Wu, T.~Su, G.~Pang, and S.-L. Chen, ``Near-{{Field Calibration}} of {{Millimeter-Wave Massive MIMO Antenna Array Using Sphere Reflectors}},'' \emph{IEEE Antennas and Wireless Propagation Letters}, pp. 1--5, 2024.

\bibitem{guoMillimeterWave3DImaging2020}
Q.~Guo, Z.~Wang, T.~Chang, and H.-L. Cui, ``Millimeter-{{Wave}} 3-{{D Imaging Testbed With MIMO Array}},'' \emph{IEEE Transactions on Microwave Theory and Techniques}, vol.~68, no.~3, pp. 1164--1174, Mar. 2020.

\bibitem{zhangMobi2Sense2022}
F.~Zhang, J.~Xiong, Z.~Chang, J.~Ma, and D.~Zhang, ``Mobi {\textsuperscript{2}} {{Sense}}: Empowering wireless sensing with mobility,'' in \emph{Proceedings of the {{Annual International Conference}} on {{Mobile Computing And Networking}} ({{MobiCom}})}, 2022, pp. 268--281.

\bibitem{zhangRFSearchSearchingUnconscious2023}
B.-B. Zhang, D.~Zhang, R.~Song, B.~Wang, Y.~Hu, and Y.~Chen, ``{{RF-Search}}: {{Searching Unconscious Victim}} in {{Smoke Scenes}} with {{RF-enabled Drone}},'' in \emph{Proceedings of the {{Annual International Conference}} on {{Mobile Computing And Networking}} ({{MobiCom}})}, 2023, pp. 1--15.

\bibitem{laiEnablingVisualRecognition2024}
H.~Lai, G.~Luo, Y.~Liu, and M.~Zhao, ``Enabling {{Visual Recognition}} at {{Radio Frequency}},'' in \emph{Proceedings of the {{Annual International Conference}} on {{Mobile Computing And Networking}} ({{MobiCom}})}, 2024, pp. 388--403.

\bibitem{tianPragmaticApproachPhase2021}
X.~Tian, Q.~Guo, Z.~Wang, T.~Chang, and H.-L. Cui, ``Pragmatic {{Approach}} to {{Phase Self-Calibration}} for {{Planar Array Millimeter-Wave MIMO Imaging}},'' \emph{IEEE Transactions on Instrumentation and Measurement}, vol.~70, pp. 1--11, 2021.

\bibitem{yuAdaptivePhasearrayCalibration2013}
J.~Yu and J.~Krolik, ``Adaptive phase-array calibration using {{MIMO}} radar clutter,'' in \emph{Proceedings of the {{IEEE Radar Conference}} ({{RadarConf}})}, Apr. 2013, pp. 1--5.

\bibitem{aumannPhasedArrayAntenna1989}
H.~Aumann, A.~Fenn, and F.~Willwerth, ``Phased array antenna calibration and pattern prediction using mutual coupling measurements,'' \emph{IEEE Transactions on Antennas and Propagation}, vol.~37, no.~7, pp. 844--850, Jul. 1989.

\bibitem{texasinstrumentsTIMmWaveRadar2018}
T.~Instruments, ``{{TI mmWave Radar}} sensor {{RF PCB Design}}, {{Manufacturing}} and {{Validation Guide}},'' 2018.

\bibitem{texasinstrumentsUsersGuideAWR1843BOOST2020}
------, ``User's {{Guide}}: {{AWR1843BOOST}} and {{IWR1843BOOST Single-Chip mmWave Sensing Solution User}}'s {{Guide}} ({{Rev}}. {{B}}),'' 2020.

\bibitem{hansenLightScatteringPlanetary1974}
J.~E. Hansen and L.~D. Travis, ``Light scattering in planetary atmospheres,'' \emph{Space Science Reviews}, vol.~16, no.~4, pp. 527--610, Oct. 1974.

\bibitem{balanisAdvancedEngineeringElectromagnetics2012}
C.~A. Balanis, \emph{Advanced {{Engineering Electromagnetics}}}, 2012.

\bibitem{liIFNetDeepImaging2024}
Y.~Li, D.~Zhang, R.~Geng, J.~Wu, Y.~Hu, Q.~Sun, and Y.~Chen, ``{{IFNet}}: {{Deep Imaging}} and {{Focusing}} for {{Handheld SAR}} with {{Millimeter-wave Signals}},'' \emph{IEEE Transactions on Mobile Computing}, pp. 1--16, 2024.

\bibitem{maMobi2SenseEnablingWireless2022}
J.~Ma, Z.~Chang, F.~Zhang, J.~Xiong, B.~Jin, and D.~Zhang, ``{{Mobi2Sense}}: Enabling wireless sensing under device motions,'' in \emph{Proceedings of the {{Annual International Conference}} on {{Mobile Computing And Networking}} ({{MobiCom}})}, 2022, pp. 766--768.

\bibitem{texasinstrumentsApplicationNoteCascade2022}
T.~Instruments, ``Application {{Note}}: {{Cascade Coherency}} and {{Phase Shifter Calibration}},'' 2022.

\bibitem{texasinstrumentsUsersGuideAWRx2020}
------, ``User's {{Guide}}: {{AWRx Cascaded Radar RF Evaluation Module}} ({{MMWCAS-RF-EVM}}),'' 2020.

\bibitem{sarabandiOptimumCornerReflectors1996}
K.~Sarabandi and T.-C. Chiu, ``Optimum corner reflectors for calibration of imaging radars,'' \emph{IEEE Transactions on Antennas and Propagation}, vol.~44, no.~10, pp. 1348--1361, Oct. 1996.

\bibitem{bleh100GHzFMCW2016}
D.~Bleh, M.~R{\"o}sch, M.~Kuri, A.~Dyck, A.~Tessmann, A.~Leuther, S.~Wagner, and O.~Ambacher, ``A 100 {{GHz FMCW MIMO}} radar system for {{3D}} image reconstruction,'' in \emph{2016 {{European Radar Conference}} ({{EuRAD}})}, Oct. 2016, pp. 37--40.

\bibitem{liuMIMORadarCalibration2018}
Y.~Liu, X.~Xu, and G.~Xu, ``{{MIMO Radar Calibration}} and {{Imagery}} for {{Near-Field Scattering Diagnosis}},'' \emph{IEEE Transactions on Aerospace and Electronic Systems}, vol.~54, no.~1, pp. 442--452, Feb. 2018.

\bibitem{crosettoPersistentScattererInterferometry2016}
M.~Crosetto, O.~Monserrat, M.~{Cuevas-Gonz{\'a}lez}, N.~Devanth{\'e}ry, and B.~Crippa, ``Persistent {{Scatterer Interferometry}}: {{A}} review,'' \emph{ISPRS Journal of Photogrammetry and Remote Sensing}, vol. 115, pp. 78--89, May 2016.

\bibitem{alvarez-narciandiShorttimeCoherentIncoherent2024}
G.~{Alvarez-Narciandi}, J.~Laviada, and F.~{Las-Heras}, ``Short-time {{Coherent Incoherent Synthetic Aperture Radar Processing}} for a {{Handheld Imaging System}},'' \emph{IEEE Transactions on Instrumentation and Measurement}, vol.~73, pp. 1--9, 2024.

\bibitem{schellbergMmSightRobustMillimeterWave2023}
J.~M. Schellberg, H.~Regmi, and S.~Sur, ``{{mmSight}}: {{Towards Robust Millimeter-Wave Imaging}} on {{Handheld Devices}},'' in \emph{Proceedings of the {{IEEE International Symposium}} on a {{World}} of {{Wireless}}, {{Mobile}} and {{Multimedia Networks}} ({{WoWMoM}})}, Jun. 2023, pp. 117--126.

\bibitem{saadatMilliCamHandheldMillimeterWave2020}
M.~S. Saadat, S.~Sur, S.~Nelakuditi, and P.~Ramanathan, ``{{MilliCam}}: {{Hand-held Millimeter-Wave Imaging}},'' in \emph{Proceedings of the {{International Conference}} on {{Computer Communications}} and {{Networks}} ({{ICCCN}})}, Aug. 2020, pp. 1--9.

\bibitem{changMSenseBoostingWireless2024}
Z.~Chang, F.~Zhang, J.~Xiong, W.~Chen, and D.~Zhang, ``{{MSense}}: {{Boosting Wireless Sensing Capability Under Motion Interference}},'' in \emph{Proceedings of the {{Annual International Conference}} on {{Mobile Computing And Networking}} ({{MobiCom}})}, Sep. 2024, pp. 1--16.

\bibitem{zhangSinglePointMultiPointReflection2024}
D.~Zhang, X.~Zhang, Y.~Xie, F.~Zhang, H.~Yang, and D.~Zhang, ``From {{Single-Point}} to {{Multi-Point Reflection Modeling}}: {{Robust Vital Signs Monitoring}} via {{mmWave Sensing}},'' \emph{IEEE Transactions on Mobile Computing}, no.~01, pp. 1--16, Aug. 2024.

\bibitem{chenCoSenseExploitingCooperative2024}
J.~Chen, D.~Zhang, G.~Zhang, H.~Wang, Q.~Sun, and Y.~Chen, ``Co-{{Sense}}: {{Exploiting Cooperative Dark Pixels}} in {{Radio Sensing}} for {{Non-Stationary Target}},'' \emph{IEEE Transactions on Mobile Computing}, pp. 1--18, 2024.

\bibitem{zhangMmRotationUnlockingVersatility2025}
D.~Zhang, X.~Zhang, Z.~Yin, Y.~Xie, H.~Wei, Z.~Chang, W.~Li, and D.~Zhang, ``{{mmRotation}}: {{Unlocking Versatility}} of a {{Single mmWave Radar}} via {{Horizontal Mobility}} and {{Azimuthal Rotation}},'' \emph{IEEE Transactions on Mobile Computing}, pp. 1--16, 2025.

\bibitem{iizukaMilliSignMmWaveBasedPassive2023}
T.~Iizuka, T.~Sasatani, T.~Nakamura, N.~Kosaka, M.~Hisada, and Y.~Kawahara, ``{{MilliSign}}: {{mmWave-Based}} passive signs for guiding {{UAVs}} in poor visibility conditions,'' in \emph{Proceedings of the {{Annual International Conference}} on {{Mobile Computing And Networking}} ({{MobiCom}})}, 2023, pp. 1--15.

\bibitem{nguyenDeepLearningFramework2023}
T.~T. Nguyen and K.-K. Nguyen, ``A {{Deep Learning Framework}} for {{Beam Selection}} and {{Power Control}} in {{Massive MIMO}} - {{Millimeter-Wave Communications}},'' \emph{IEEE Transactions on Mobile Computing}, vol.~22, no.~8, pp. 4374--4387, Aug. 2023.

\end{thebibliography}

\vfill

\end{document}